\documentclass[preprint,12pt]{elsarticle}

\usepackage{booktabs}
\usepackage{amsmath}
\usepackage{amsfonts}
\usepackage{amssymb}
\usepackage{array}
\usepackage{lscape}
\usepackage{geometry}
\usepackage{longtable}
\usepackage{float}
\usepackage[hyphens]{url}
\biboptions{sort&compress}

\usepackage{nomencl}
\makenomenclature
\usepackage{etoolbox}
\renewcommand\nomgroup[1]{%
  \item[\bfseries
  \ifstrequal{#1}{S}{Sets}{%
  \ifstrequal{#1}{P}{Parameters}{%
  \ifstrequal{#1}{V}{Variables}{%
  \ifstrequal{#1}{A}{Abbreviations}{}}}}%
]}

\usepackage{amssymb}
\usepackage{placeins}

\usepackage{siunitx}

\journal{Smart Energy}

\begin{document}

\begin{frontmatter}



\title{Solar prosumage under different pricing regimes: Interactions with the transmission grid}

\author[diw]{Dana Kirchem \corref{cor1}}
\ead{dkirchem@diw.de}
\cortext[cor1]{corresponding author}

\author[tub]{Mario Kendziorski}

\author[tub]{Enno Wiebrow}
\ead{ewi@wip.tu-berlin.de}

\author[diw]{Wolf-Peter Schill}
\ead{wschill@diw.de}

\author[diw]{Claudia Kemfert}
\ead{ckemfert@diw.de}

\author[tub]{Christian von Hirschhausen}
\ead{cvh@wip.tu-berlin.de}

\affiliation[diw]{organization={DIW Berlin},
            addressline={Mohrenstraße 58}, 
            city={Berlin},
            postcode={10117}, 
            country={Germany}}

\affiliation[tub]{organization={TU Berlin},
            addressline={Straße des 17.~Juni 135}, 
            city={Berlin},
            postcode={10623}, 
            country={Germany}}

\begin{abstract} 
Solar prosumers, residential electricity consumers equipped with photovoltaic (PV) systems and battery storage, are transforming electricity markets. Their interactions with the transmission grid under varying tariff designs are not yet fully understood. We explore the influence of different pricing regimes on prosumer investment and dispatch decisions and their subsequent impact on the transmission grid. Using an integrated modeling approach that combines two open-source dispatch, investment and grid models, we simulate prosumage behavior in Germany's electricity market under real-time pricing or time-invariant pricing, as well as under zonal or nodal pricing. Our findings show that zonal pricing favors prosumer investments, while time-invariant pricing rather hinders it. In comparison, regional solar availability emerges as a larger driver for rooftop PV investments. The impact of prosumer strategies on grid congestion remains limited within the scope of our model-setup, in which home batteries cannot be used for energy arbitrage.
\end{abstract}




\end{frontmatter}



\newpage

\section{Introduction}\label{sec: introduction}
In many electricity markets, grid-connected consumers can minimize their energy bills by using rooftop PV systems and battery storage to self-generate parts of their electricity consumption. This phenomenon, which has been referred to as solar prosumage \cite{von_hirschhausen_prosumage_2017,schill_prosumage_2017}, is on the rise in electricity markets around the world. In Germany, solar prosumage has been growing significantly in recent years, and projections indicate a continued upward trajectory as the overall PV capacity is planned to quadruple to 400~GW, with rooftop systems accounting for around half of it \cite{schmidt2024}. 

Solar prosumage has also become more relevant in other countries, such as Australia \cite{say2018,say_degrees_2020}, France \cite{roulot2018} and the United States \cite{tervo2018}. While solar availability remains one of the key determinants of economic viability, such that in countries like Ireland, prosumer incentives are not sufficient for market uptake in the current policy environment \cite{bertsch2017}, a study for Spain also demonstrates that even with high solar availability, policy design is crucial for the uptake in prosumage \cite{prol2017}.

The decision-making of prosumers still largely remains independent of wholesale market price signals, while retail tariffs, feed-in tariffs and PV investment costs play a pivotal role in shaping their investment and dispatch decisions \cite{gunther_prosumage_2021}. Among these factors, lower feed-in tariffs can reduce PV investments, while low variable retail tariff parts can reduce PV self-generation and battery investments \cite{gunther_prosumage_2021}. Network charges and levies are also relevant for prosumer decisions \cite{thomsen2021}. Regarding storage investment costs, home batteries have to be very cheap for high self-consumption to be economically viable for prosumers \cite{green2017}. 

While the effects of solar prosumage on other generation and storage capacities in the power sector have been previously studied \cite{schill_prosumage_2017}, their interactions with the electricity grid are not yet fully understood. On distribution grid level, prosumage can lead to additional distribution grid stress, which can be mitigated by policies such as a maximum grid feed-in of rooftop PV \cite{neetzow2019}. Furthermore, grid stress can be reduced when prosumage systems are operated with a load and solar availability forecast rather than with the premise of self-consumption maximization \cite{moshovel2015}.

Evidence regarding the effects of prosumage on the transmission grid is more ambiguous though. Prosumage may help defer transmission grid investments by reducing peak PV feed-in levels \cite{schill_prosumage_2017}. This can lead to lower transmission network usage and losses, particularly when prosumage smoothes PV peak feed-in. \cite{chen2023} find that on the one hand, the additional renewable energy provided by prosumers can cause transmission charges to decrease. On the other hand, strategic behavior of prosumers could also increase transmission charges. 

In this context, the implications of electricity tariff design for investments of rooftop PV and home batteries have not yet been connected to transmission grid effects in the current literature Understanding these interactions is relevant for efficient grid planning and operation and for developing market structures that accommodate the growing trend of decentralized self-generation. This study aims to bridge this gap by analyzing how different pricing regimes influence investments in solar photovoltaic capacities and home batteries, and how these in turn impact the transmission grid. To achieve this, we combine two open-source models to quantitatively assess the potential impacts. First, the economic dispatch model ELMOD generates wholesale price time series for future scenarios in 2030 in Germany. These electricity prices serve as the basis for investment decisions into rooftop PV and battery storage using a prosumage module of the model DIETER, which takes into account different tariff design assumptions. Finally, the impacts of these investment decisions on the transmission grid are evaluated with ELMOD. This multi-stage approach allows us to provide a comprehensive analysis of how different pricing regimes influence prosumer behaviour and, consequently, the transmission grid.

\section{Methods}
\subsection{The Model ELMOD}
ELMOD \cite{leuthold_elmod_2008, egerer_2016, weibezahn_illustrating_2019} is a comprehensive multi-step electricity market and transmission grid model, incorporating both zonal market clearing with subsequent redispatch and nodal pricing representation. By integrating congestion management, ELMOD aims to compare different market designs and their impact on market outcomes. The tool leverages a DC load flow approach to model power flows and analyze effects on the grid. 


\subsubsection{Day-Ahead model}
In the day-ahead model, ELMOD minimizes the marginal costs of electricity generation and storage as well as the costs of unit curtailment according to the objective function in Equation \ref{eq:ELMOD_DA_OF}.

\begin{equation}\label{eq:ELMOD_DA_OF}
\begin{aligned}
    \min \quad & \sum_{t}^T \sum_{p}^{P} mc_{p,t} \cdot GEN_{p,t} 
     + c^{curt} \cdot \sum_{t}^{T} \sum_{z}^{Z}CU_{z,t} 
    & + \sum_t^T \sum_{s}^{S} mc_{s,t} \cdot GEN_{s,t}
\end{aligned}
\end{equation}

Based on the selected market configuration, either the \textit{Zonal Market Balance} (Equation \ref{eq:ELMOD_DA_zonal}) with exchange constraints or the \textit{Nodal Market Balance} (Equation \ref{eq:ELMOD_DA_nodal}) with DC load flow constraints is implemented. Note that the difference lies in taking into account the net exchange of the zone $EX_{z,t}^{net}$ with its neighboring zones in Equation \ref{eq:ELMOD_DA_zonal}, while taking into account the injection $INJ_{n,t}$ at each node in Equation \ref{eq:ELMOD_DA_nodal}. 

\begin{equation}\label{eq:ELMOD_DA_zonal}
\begin{aligned}
& \sum_{p \ in \ z}^{P} GEN_{p,t} 
+  \sum_{s \ in \ z}^{S} (GEN_{s,t} - CHARGE_{s,t}) 
  +  EX_{z,t}^{net}
  -   CU_{z,t} \\
& = \sum_{n \ in \ z}^{N} load_{n,t} 
 - LL_{z,t}, & \forall \ z \in Z, t \in T
\end{aligned}
\end{equation}

\begin{equation}\label{eq:ELMOD_DA_nodal}
 \begin{aligned}
    &\sum_{p \ in \ n}^P GEN_{p,t} +
    \sum_{s \ in \ n}^S (GEN_{s,t} -CHARGE_{s,t})
    &+ INJ_{n,t} - CU_{n,t}\\ = &load_{n,t} - LL_{n,t} & \forall \ n \in N, t \in T
\end{aligned}   
\end{equation}


\subsubsection{Redispatch model}
In the redispatch phase of ELMOD, redispatch costs are minimized according to the objective function in Equation \ref{eq:ELMOD_RD_OF}. Total redispatch costs comprise of the redispatch costs of generating units $P$ and storage units $S$, as well as the curtailment costs of units after the redispatch.

\begin{equation}\label{eq:ELMOD_RD_OF}
\begin{aligned}
\min \quad & \sum_t^T\sum_{p}^P c^{redisp} \cdot (GEN_{p,t}^{up} + GEN_{p,t}^{down}) 
 +  \sum_t^T\sum_{p}^P c^{curt} \cdot (CU_{p,t}^{redisp} - cu_{p,t}) \\
& + \sum_t^T\sum_{s}^S c^{redisp} \cdot (GEN_{s,t}^{up} + GEN_{s,t}^{down})\\
& + \sum_t^T\sum_{s}^S c^{redisp} \cdot (CHARGE_{s,t}^{up} + CHARGE_{s,t}^{down})
\end{aligned}
\end{equation}

The redispatch market balance (Equation \ref{eq:ELMOD_RD_Balance}) resembles the nodal market balance of the day-ahead model, but with the updated unit and storage operation after the redispatch.

\begin{equation}\label{eq:ELMOD_RD_Balance}
 \begin{aligned}
    &\sum_{p \ in \ n}^P GEN_{p,t}^{redisp} +
    \sum_{s \ in \ n }^S (GEN_{s,t}^{redisp} -CHARGE_{s,t}^{redisp}) \\
    &+ INJ_{n,t} - CU_{n,t}
     + \sum_{prs \ in \ n}^{PRS} GEN_{prs,t}\\ & = load_{n,t} - LL_{n,t} & \forall \ n \in N, t \in T
\end{aligned}   
\end{equation}

\subsection{The Model DIETER}
DIETER (Dispatch and Investment Evaluation Tool with Endogenous Renewables) is an open-source power sector model \cite{gaete2021} which can incorporate different energy sectors, such as mobility, heating or hydrogen production \cite{kirchem2023hydrogen,roth2024,gaetemorales2024,gueret2024}.\footnote{The model code for this study is available here: \url{https://gitlab.com/diw-evu/projects/prosumage-lp}.} In its main version, DIETER is a linear program that determines the least-cost capacity investment and dispatch decisions for various electricity generation and storage technologies. 

In this analysis, we do not use the full model, but only a prosumage module, leaning on \cite{gunther_prosumage_2021}. This model does not include the full power sector, but relies on external wholesale electricity prices. Here, prosumers minimize their overall electricity bill according to equation \ref{eq:dieter_OF}. They optimize their investments into PV capacity $INVEST_{pv}$ and home batteries $INVEST_{s^e}$, their direct self-consumption $G^{self}_{pv,t}$, their self-consumption via storage $GEN_{pv,t}$, their electricity usage from the grid $F_t$ and the grid feed-in $GEN^{grid}_{pv,t}$, subject to the electricity price (RTP or time-invariant) with perfect foresight and subject to a constant feed-in tariff $t^{feed}$. Prosumers satisfy their electricity demand by grid consumption, battery storage outflows and self consumption (see equation \ref{eq:dieter_balance}). Household batteries can only be charged with energy from rooftop PV. 

Regarding PV investments, we assume that the extension potential solely involves private residential rooftops. We project that every suitable household in Germany will not only have a solar rooftop but also a corresponding solar battery. The investments in solar rooftop PV and batteries are anticipated to substitute the PV and battery capacities previously projected in the price forecast.

\begin{equation}\label{eq:dieter_OF}
\begin{aligned}
    \min \quad & (c^{invest}_{pv} + c^{fix}_{pv}) \cdot INVEST_{pv} + c^{invest}_{s^e} \cdot INVEST_{s^e}\\
    & + c^{invest}_{s^p} \cdot INVEST_{s^p} + c^{fix}_{s}/2 \cdot (INVEST_{s^p} + INVEST_{s^e})\\
    & + \sum_t^T ( mc_{pv} \cdot GEN_{pv,t} + mc_{s} \cdot (CHARGE_{s,t} + GEN_{s,t}) \\
    & + (t^{fix} + t^{var} + binary^{nodal} \cdot rtp^{nodal}_t + binary^{zonal} \cdot rtp^{zonal}_t) \cdot F_t\\
    & - (t^{feed} + binary^{nodal} \cdot rtp^{nodal}_t + binary^{zonal} \cdot rtp^{zonal}_h) \cdot GEN^{grid}_{pv,t} 
        )
\end{aligned}
\end{equation}

\begin{equation}\label{eq:dieter_balance}
demand_{t} \leq F_t + GEN_{s,t} + GEN^{self}_{pv,t}
\end{equation}

\subsection{Model linkage}
\begin{figure}
    \centering
    \includegraphics[width=0.5\textwidth]{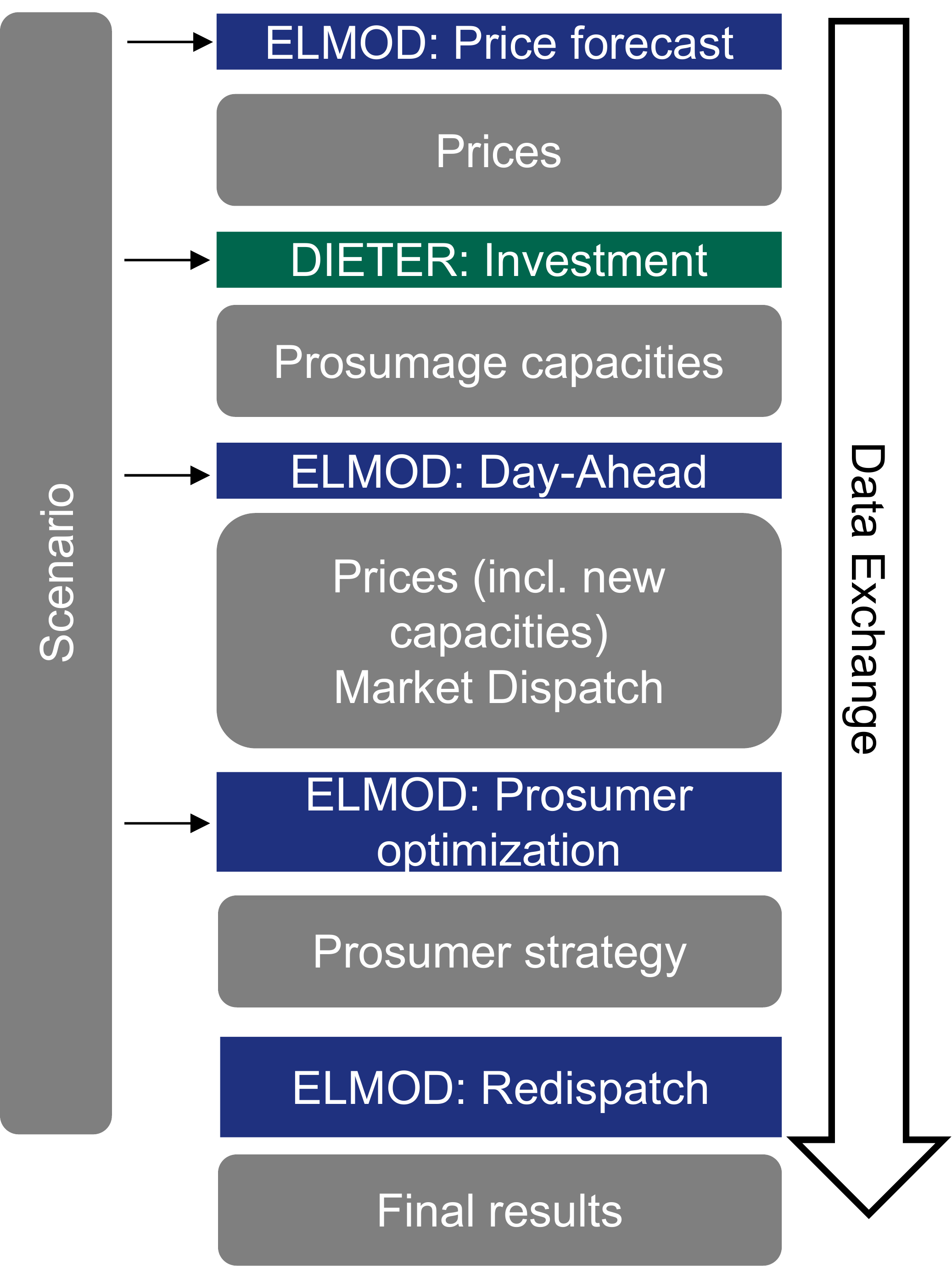}
    \caption{Illustration of the model interaction}
    \label{fig:models}
\end{figure}

To investigate the interactions between solar prosumage and the transmission grid, we integrate the two open-source models ELMOD and DIETER. Our approach focuses on leveraging ELMOD for price forecasting and using the prosumage module of the model DIETER \cite{gunther_prosumage_2021} to model investment and dispatch decisions (illustrated in figure \ref{fig:models}).

\paragraph{Step 1: Price Forecast with ELMOD}
First, ELMOD performs a price forecast for the parameterized scenario. This model clears either a zonal market (where Germany functions as a single price zone) or a nodal market (where each network node has its unique price) depending on the scenario. The projected wholesale electricity prices are subsequently utilized in the ensuing step.

\paragraph{Step 2: Prosumage Investment Modeling with DIETER}
Next, DIETER uses the wholesale price forecasts from ELMOD to model investments in rooftop PV and battery capacity. These investments represent cost-minimizing prosumage behavior and are influenced by different electricity pricing scenarios:

\begin{itemize}
    \item Time-Invariant Prices: Simulates a fixed tariff that remains constant, as experienced by most retail customers.
    \item Real-Time Pricing: Reflects day-ahead market price variations in real-time, providing a dynamic pricing environment.
\end{itemize}

Note that the feed-in tariff remains the same across scenarios.

\paragraph{Step 3: Day-Ahead Market Simulation with Updated Capacities}
The investment decisions from DIETER are taken as updated rooftop PV and home battery capacities for subsequent simulations.

\paragraph{Step 4: Prosumer Optimization}
Depending on the market pricing scenario (real-time vs. time-invariant, zonal vs. nodal), prosumers tailor their battery dispatch strategies to enable solar self-consumption patterns that minimize their energy bills. 

\paragraph{Step 5: Intraday and Redispatch Stage}
Finally, the prosumers feed-in to the grid and their demand from the grid are used as inputs for the intraday and redispatch stage. Here, any potential imbalances in generation or grid congestion are resolved through redispatch actions.\\ 

\section{Input data and scenarios}

We parameterize our model analysis for a German case study, leaning on the Scenario C of Germany's grid development plan (Netzentwicklungsplan, NEP) \cite{bnetza_genehmigung_2018} for the year 2030. This scenario represents an ambitious outlook in terms of prosumage and flexibility technologies. It envisages a share of 65\% renewable energy sources in gross electricity consumption, which is likely to be achieved well before the year 2030 and thus can be interpreted as a near-term future scenario.

\begin{table}[h]
    \centering
    \begin{tabular}{l c}
        \toprule
        \textbf{Technology} & \textbf{Installed capacity [GW]} \\
        \midrule
        Hard coal          & 8.1 \\
        Lignite            & 9 \\
        Gas                & 33.4 \\
        Other conventionals & 5 \\
        Biomass            & 8.5 \\
        Hydro              & 10.4 \\
        Wind onshore       & 85.5 \\
        Wind offshore      & 18.3 \\
        Ground-mounted PV (solar park)         & 31.2 \\
        Rooftop PV      & Endogenous \\
        Home battery       & Endogenous \\
        \bottomrule
    \end{tabular}
    \caption{Installed Capacity by Technology.}
\end{table}

Our data set encompasses 640 transmission grid nodes and over 1500 transmission lines, based on a data set generated with the open-source Power Market Tool (POMATO) \cite{weinhold_power_2021}. Renewable time series with high spatial and temporal resolution are generated at the nodal level using the atlite tool \cite{hofmann_atlite_2021} which enables accurate simulation of renewable generation profiles, particularly for wind and solar power. Regionalization of wind onshore and ground-mounted solar PV (solar parks) capacities is performed based on regional extension potentials.

The maximum installable PV capacity for households within each NUTS3 zone is calculated based on the extension potential data cited by \cite{ebner_regionalized_2019}, with a set capacity limit of \SI{10}{\kilo\watt} per household. Households have an average electricity demand of 3500~kWh per year. In order to set a maximum installable PV capacity at every node, we divide the total household electricity demand at each node by 3500 to obtain the number of representative households. This is then multiplied by 10~kW, which gives the maximum installable PV capacity at every node.

Market simulations are conducted for one year at an hourly resolution. Two market designs are simulated: zonal pricing with redispatch (the current market design) and nodal pricing. The current market design represents the existing structure where wholesale electricity prices vary hourly, but are uniform across the market zone and congestion is managed via redispatch. Nodal pricing involves location-specific hourly wholesale market prices that reflect the local supply-demand balance considering transmission grid congestion.

Regarding tariff schemes, we further distinguish between time-invariant and real-time pricing (RTP). In combination, this results in four different pricing regimes: zonal time-invariant, zonal RTP, nodal time-invariant and nodal RTP. The feed-in tariff for household-produced electricity is set at 0.06~Euro per kWh. Real-time prices are obtained from the price forecast of ELMOD, while time-invariant prices are calculated as a demand-weighted average of this price time series over the full year. The electricity tariff comprises a variable non-energy payment of 0.25~Euro per kWh and no upfront payment.

\section{Results}
\subsection{Investment decisions}
\begin{figure}
\centerline{\includegraphics[width=\columnwidth]{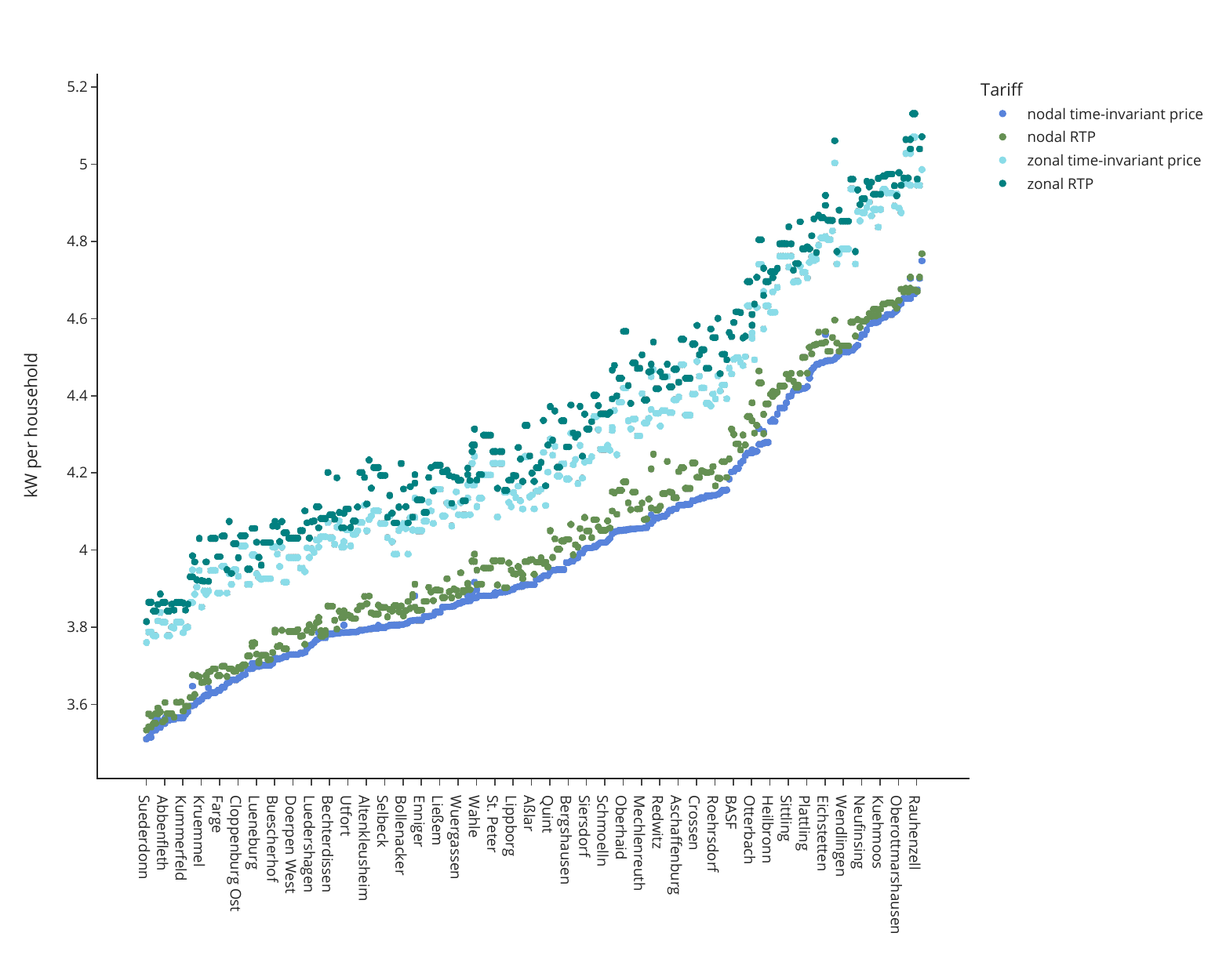}}
\caption{Investments in rooftop PV per (representative) household}
\label{fig:pv_per_hh}
\end{figure}

Investments into rooftop PV vary between 3.51~kW per household in the North of Germany and 5.13~kW per household in the South (see Figure~\ref{fig:pv_per_hh}). The increasing investments from northern to southern regions in Germany are due to higher solar irradiation, which translates into higher full-load hours of solar PV. On average, we observe that zonal pricing leads to substantially higher rooftop PV investments than nodal pricing. This is driven by wholesale price differences. Prices in the zonal scenario (0.35~EUR/kWh) are on average around twice as high than in the nodal scenario (0.17~EUR/kWh), which makes higher levels of prosumage more profitable. Real-time pricing further leads to slightly higher rooftop PV adoption as compared to time-invariant pricing. 
This is because real-time pricing offers the opportunity for prosumers to align the electricity self-consumption with the electricity wholesale market and to substitute grid consumption in high-price hours to a larger extent. The highest PV investments are achieved with a combination of zonal and real-time pricing.   

\begin{figure}
\centerline{\includegraphics[width=\columnwidth]{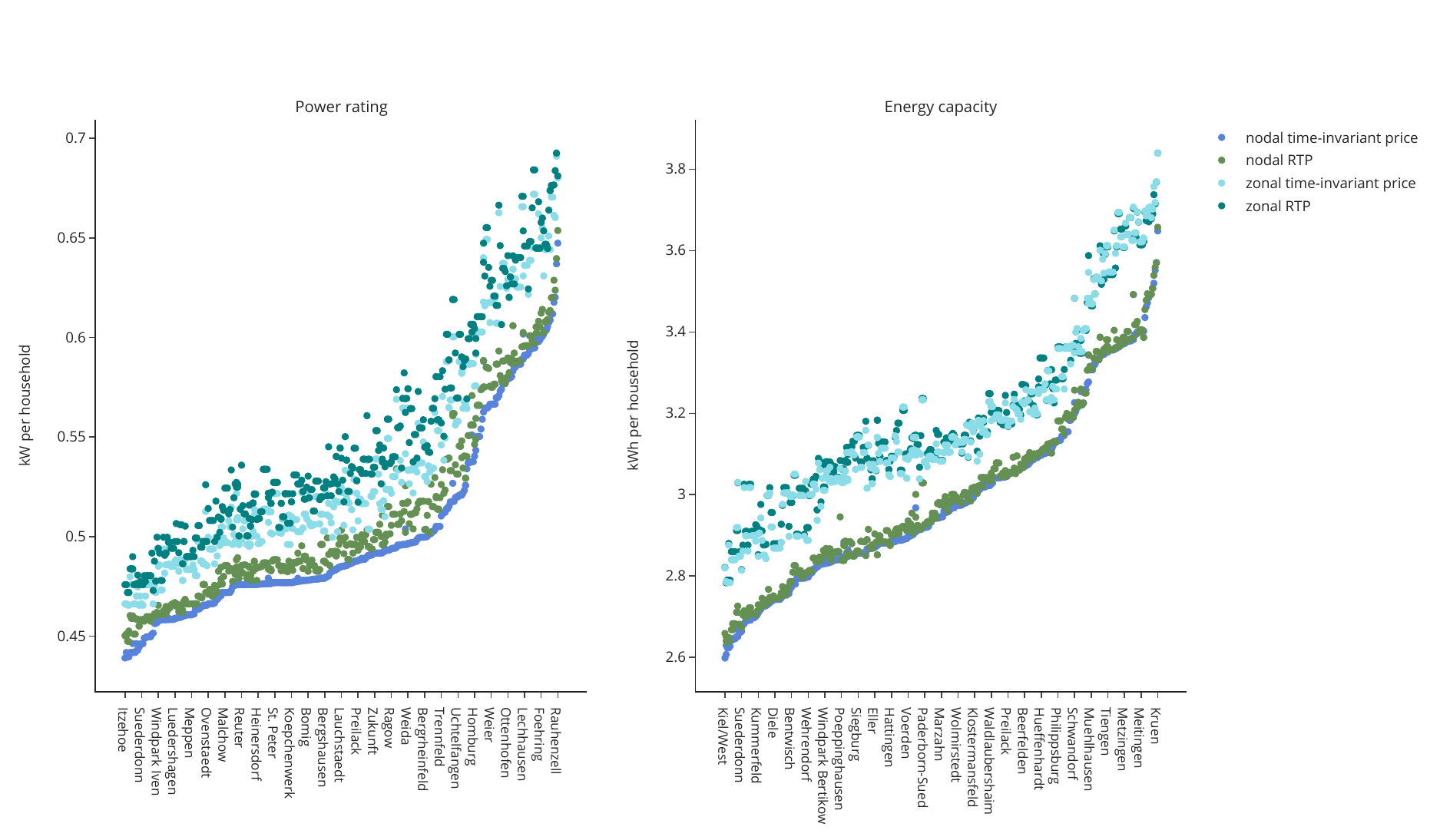}}
\caption{Investments in home batteries per (representative) household}
\label{fig:battery_per_hh}
\end{figure}

We observe similar effects for home battery investments. These tend to be the higher in the scenario with zonal pricing, which makes self-consumption more profitable because of higher average wholesale prices. Additionally, consumers can avoid high-price periods by storage discharging if they are subjected to real-time pricing, thus they tend to invest more into home batteries. Power ratings of home batteries range from 0.4~kW to 0.7~kW, while energy capacity is between 2.6~kWh and 3.8~kWh (see Figure~\ref{fig:battery_per_hh}). In terms of storage energy capacity, the effect of real-time pricing on installed capacity is smaller than for storage power rating or PV installations. Zonal pricing still has a positive effect on installed energy capacity. In general, we do not observe a large difference in investments based on the different pricing regimes, while the location of the node and thus the availability factor of solar energy plays a much larger role.  

\subsection{Dispatch decisions}
Feeding the investment decisions back into ELMOD reveals how different pricing schemes influence prosumer dispatch strategies. The analysis compares time-invariant and real-time pricing.\\

Under time-invariant pricing, the main objective is to use stored energy immediately to minimize losses from self-discharge, given that a very small self-discharge rate (lower than 0.01\% per hour) is implemented. In contrast, when a temporal price signal is present, such as with real-time pricing, the determining factor becomes the price of grid electricity. Here, the stored electricity is not discharged immediately when domestic demand exceeds PV supply, but at least partly in later periods when electricity prices are high, e.g., in early morning hours.

\begin{figure}
    \centering
    \includegraphics[width=\textwidth]{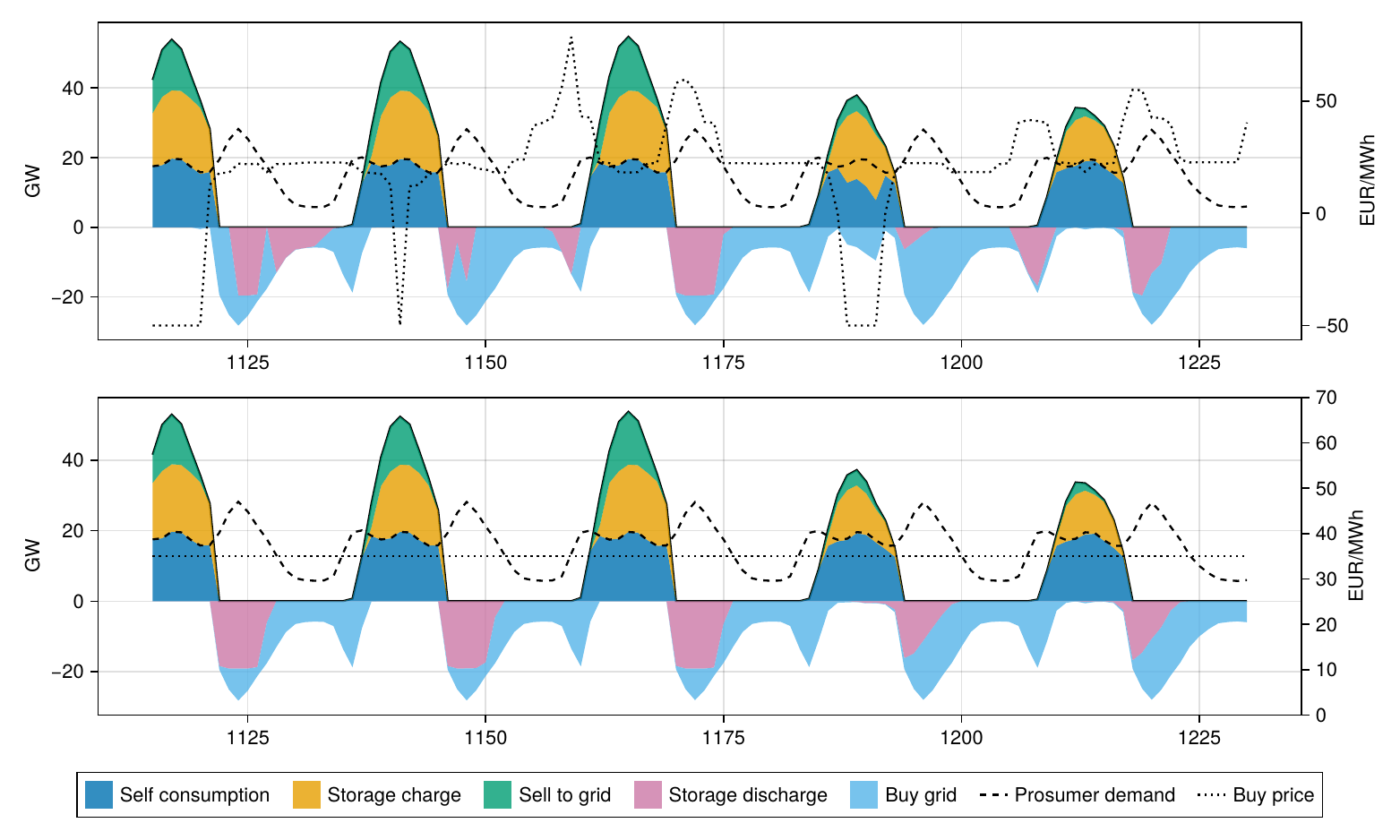}
    \caption{Optimal prosumer dispatch decisions for real-time pricing (upper panel) and time-invariant pricing (lower panel) for five selected days.}
    \label{fig:prs_strategy}
\end{figure}

Figure \ref{fig:prs_strategy} compares the prosumer strategy under both pricing schemes for an exemplary time period. In the case of time-invariant pricing, prosumers consume stored energy immediately to avoid losses from self-discharge while prosumers with real-time pricing discharge their batteries to minimize the costs of grid electricity consumption. They prioritize purchasing electricity during low-price periods and consuming stored energy during high-price periods, creating a more sophisticated dispatch strategy compared to time-invariant pricing.

The pricing scheme hardly affects the amount of energy that prosumers feed in to the grid, since there is no incentive to alter the strategy of maximizing energy charging with the available battery capacity until it is completely charged. However, the demand pattern for electricity from the network changes significantly if prosumers have a real-time pricing scheme.

\begin{figure}
    \centering
    \includegraphics[width=\textwidth]{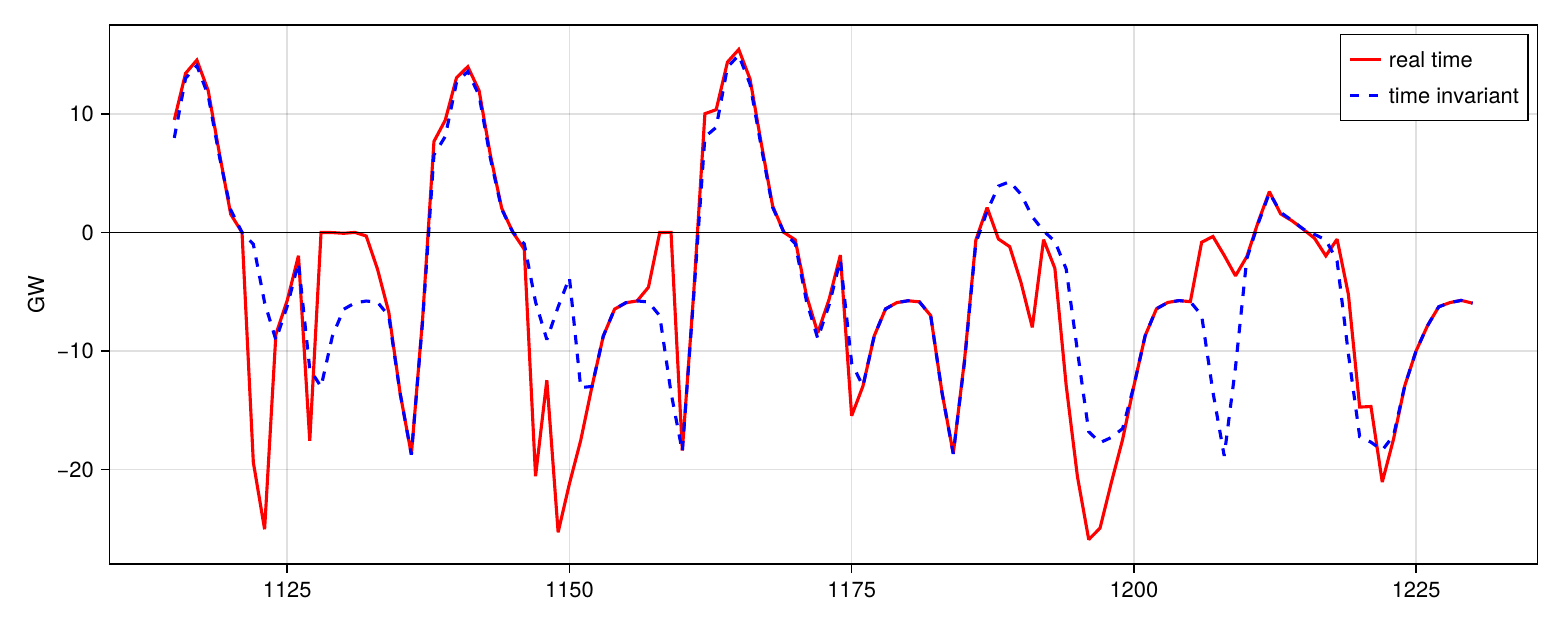}
    \caption{Resulting net input of prosumers to the grid.}
    \label{fig:prs_netinput}
\end{figure}

Figure \ref{fig:prs_netinput} shows the net grid interaction of prosumers under both time-invariant pricing and real-time pricing. Under real-time pricing, grid consumption peaks are higher because prosumers buy more electricity during low-price periods. In addition, prosumers avoid grid consumption by consuming previously stored energy during high-price hours. This behavior is particularly noticeable when prices are relatively low in the evening and higher the next morning. In such cases, prosumers avoid buying electricity during high-price periods by utilizing stored energy from the previous day.

\begin{figure}
    \centering
    \includegraphics[width=\textwidth]{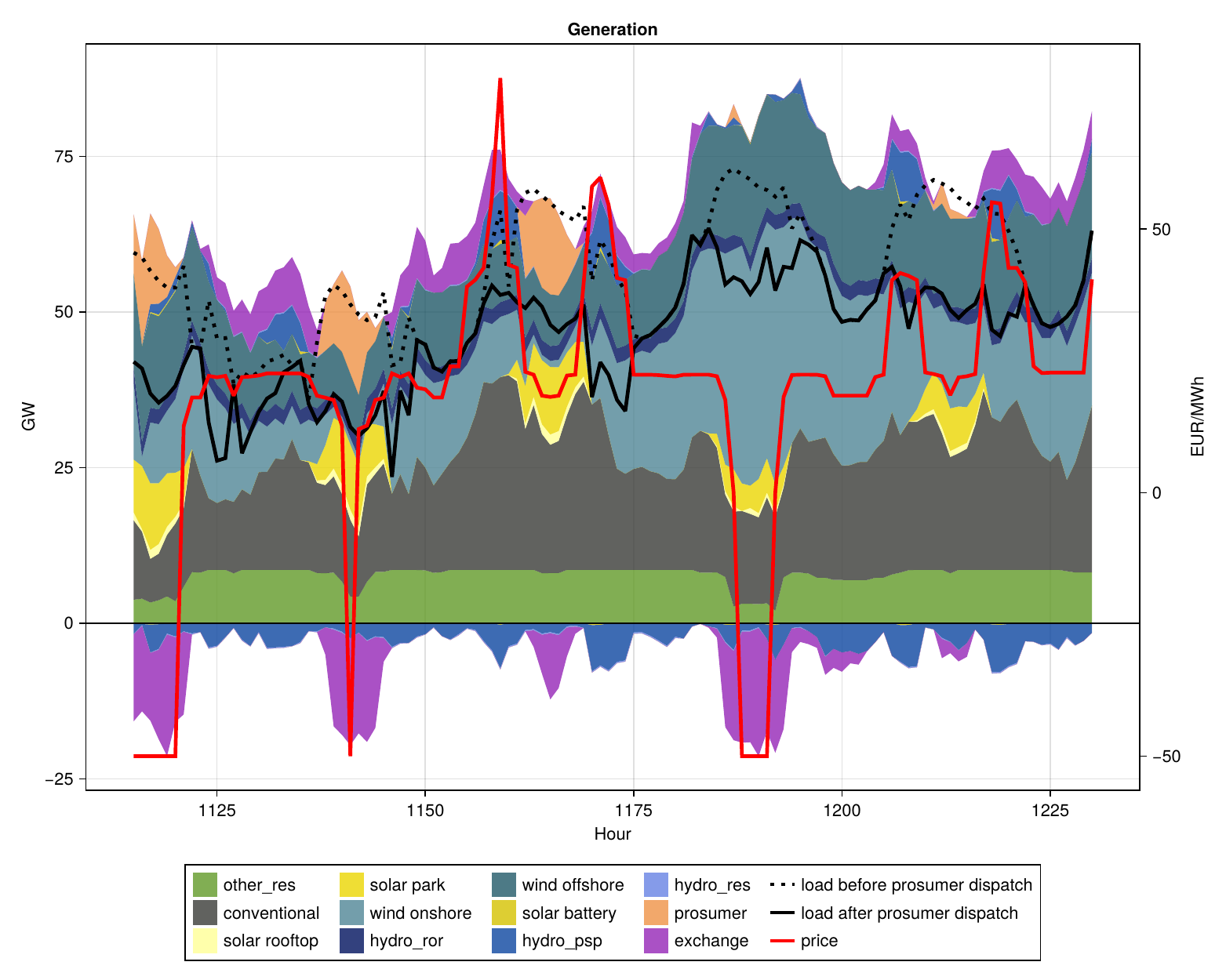}
    \caption{Example dispatch in summer week}
    \label{fig:dispatch_rt_zonal}
\end{figure}

Figure \ref{fig:dispatch_rt_zonal} depicts the use of all generation and storage technologies for the real-time pricing scenario in the same time period. The generation from prosumers peaks in the middle of the day when solar PV generation is at its highest. During these hours, electricity is often exported to neighboring markets. The system load after the prosumer dispatch decision (solid black line in Figure \ref{fig:dispatch_rt_zonal}) is lowered in two situations:

\begin{enumerate}
    \item Low-price periods: Prosumers satisfy their own demand and feed excess PV electricity to the grid. Thus, the load is reduced during these hours.
    \item High-price periods: The total load is mostly lowered during times with higher prices because prosumers discharge their batteries and use less electricity from the grid, resulting in a smoother load curve with fewer peaks.
\end{enumerate}
    
These results demonstrate that real-time pricing leads to more dynamic prosumer behavior and a smoother demand curve compared to time-invariant pricing because of an optimized operation of the home battery. Note that this is true even without further grid interactions of prosumer batteries, and while keeping a time-invariant feed-in tariff.

\begin{figure}
    \centering
    \includegraphics[width=0.8\textwidth]{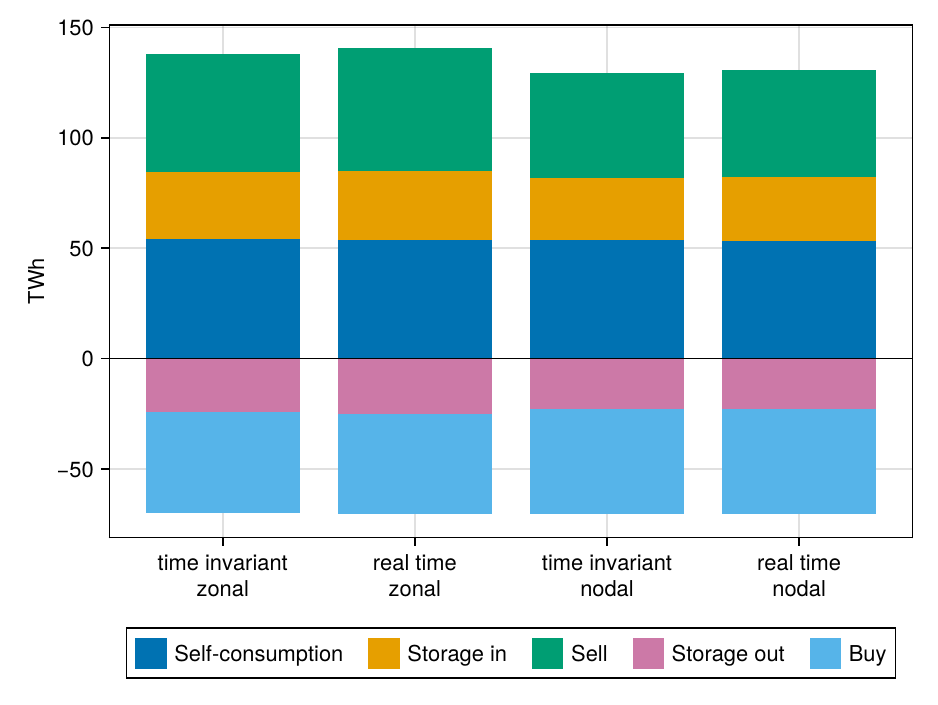}
    \caption{Aggregated values for prosumer dispatch.}
    \label{fig:prs_summary}
\end{figure}

Although real-time pricing leads to more dynamic prosumer behavior and refined battery dispatch strategies, the annual difference between the scenarios remains small. Prosumers aim to minimize costs by shifting battery discharge to more advantageous hours, but this optimization does not significantly alter the total amount of self-consumption, stored energy, or electricity exchanged with the grid. To some extent, this finding might also be driven by our model sequence, according to which the prosumer investments are fixed in step 2 of the model linkage. Figure \ref{fig:prs_summary} shows that the yearly aggregated values for self-consumption, stored energy, and electricity sold to and bought from the grid remain almost identical across the two pricing schemes.

The annual generation of other market participants also remains relatively stable across different pricing schemes. Table \ref{tab:generation} provides a comparative overview of total generation from various sources besides prosumers under time-invariant and real-time pricing scenarios.

\begin{table}
\begin{tabular}{@{}lrrrr@{}}
\toprule
plant type   & real time & time-invariant& real time & time-invariant \\ 
&zonal&zonal&nodal&nodal\\\midrule
other\_res    & 57.6            & 57.7                 & 62.2            & 62.2                 \\
conventional  & 172.7           & 173.7                & 162.3           & 163.0                \\
solar rooftop & 3.6             & 3.7                  & 4.2             & 4.2                  \\
Ground-mounted PV    & 23.7            & 24.0                 & 27.2            & 27.3                 \\
(solar park)&&&&\\
solar battery & 0.8             & 0.7                  & 0.5             & 0.5                  \\
wind onshore  & 139.6           & 140.0                & 135.2           & 135.3                \\
wind offshore & 79.0            & 79.3                 & 76.3            & 76.2                 \\
hydro\_ror    & 20.8            & 20.9                 & 21.7            & 21.7                 \\
(run-of-river)&&&&\\
hydro\_psp    & 19.9            & 19.8                 & 11.8            & 11.7                 \\
(pumped storage)&&&&\\\bottomrule
\end{tabular}
    \caption{Yearly generation by plant type and scenario (TWh).}
    \label{tab:generation}
\end{table}

\subsection{Transmission grid}
Table \ref{tab:redispatch} shows the redispatch after the prosumer battery dispatch in terms of generation increase and decrease to balance the system. The model setup balances mismatches in generation caused by temporal deviations due to prosumers' dispatch decisions and network congestion in the same step, meaning that these two effects cannot be separated clearly.

\begin{table}
    \centering
    \begin{tabular}{@{}lll@{}}
    \toprule
    scenario             & increase & decrease \\ \midrule
    real time zonal      & 156.7    & 31.8     \\
    time invariant zonal & 157.2    & 30.7     \\
    \bottomrule
    \end{tabular}

    \caption{Total annual generation increase and decrease after the prosumer decision stage (TWh).}
    \label{tab:redispatch}
\end{table}

\begin{figure}
    \centering
    \includegraphics[width=0.95\textwidth]{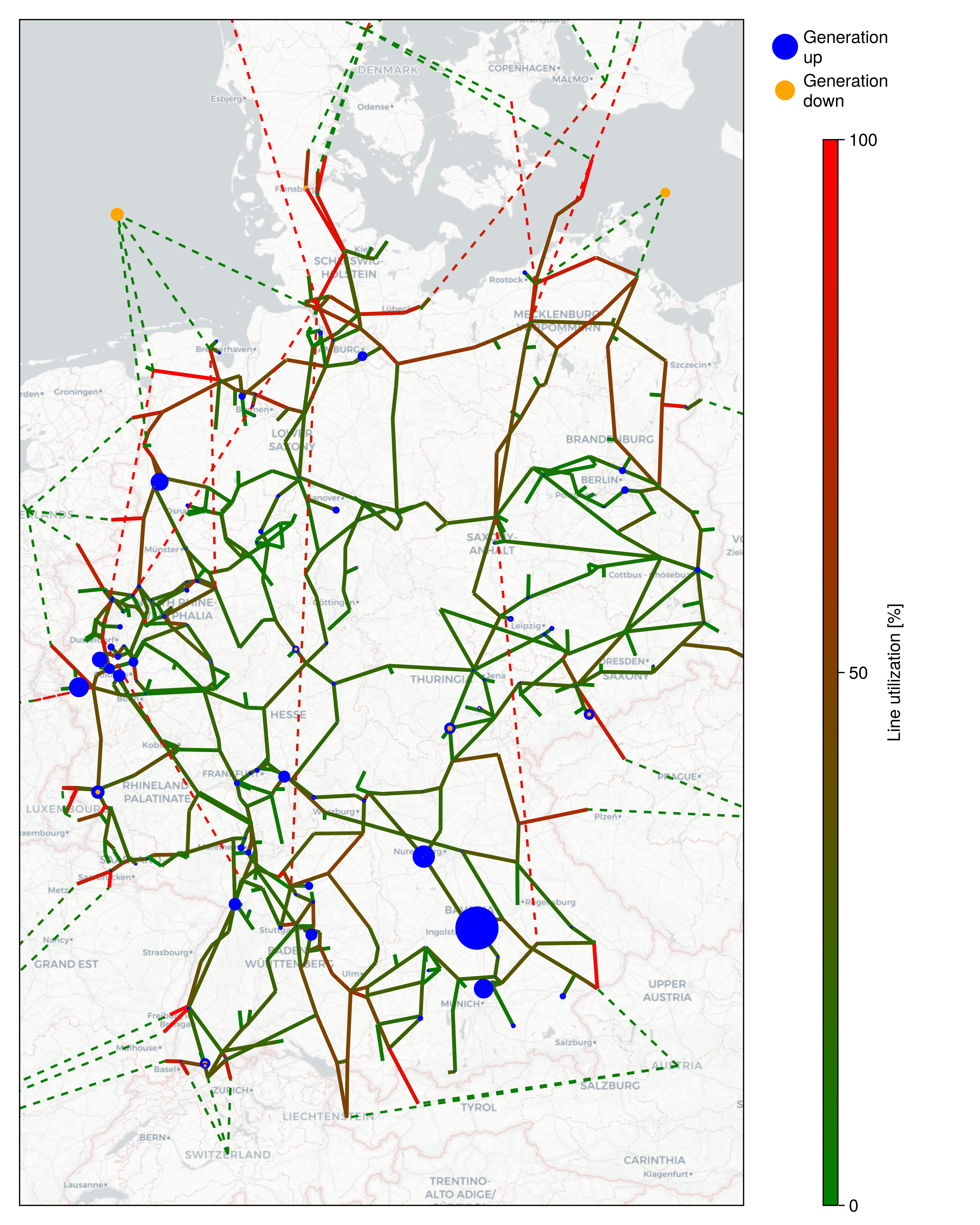}
    \caption{Average line utlization and redispatch measures for real time pricing in zonal market.}
    \label{fig:util_line_map}
\end{figure}

Figure \ref{fig:util_line_map} depicts the average line utilization and corresponding generation increases and decreases. The most utilized lines are almost entirely those connected to other market areas, indicating that the main driver for network congestion is the exchange with neighboring markets. This observation leads to the conclusion that the adaptation of generation is to a large extent a redispatch measure not caused by prosumers' decisions but rather by the specific model setup and dataset. A comparison of relative values between the scenarios shows that we cannot observe a significant impact on the adaptation measures.\\

\subsection{Limitations}
Our results provide insights into the relationship between pricing regimes and solar prosumage, and on its limited transmission grid impacts in the German scenario modeled here. Yet, several limitations must be acknowledged and discussed. First, the model setup assumes that prosumer behavior is not anticipated in the day-ahead market (step 1). Instead, the market is cleared as if prosumer battery dispatch could be optimized in the most cost-effective way from a welfare perspective. In reality, with a significant share of generation and demand coming from prosumers, other market participants would anticipate their behavior. This anticipation would likely lead to different market prices as other market participants would adjust their bidding strategies in response to the expected prosumer dispatch patterns.\\ 

Second, the analysis does not consider flexibility options for households within the electricity market other than home batteries. With the increasing penetration of electric vehicles and heat pumps, there will be more opportunities for prosumers to optimize their consumption strategies. Other flexible ways to react to a real time price signal such as smart home automation or microgrid participation could significantly alter their behavior. This could lead to higher differences between the pricing schemes and more dynamic effects on the transmission grid. Likewise, we do not allow any further grid interactions of prosumer batteries. While this reflects the status quo in Germany, it appears desirable that PV-batteries would also provide their flexibility to the overall electricity sector in the future, for example by engaging in energy arbitrage guided by wholesale prices. Yet, this is likely to lead to substantially different transmission grid impacts than modeled here under zonal pricing, as storage charging and discharging might even aggravate existing grid congestion.\\

Third, the model assumes that prosumers have perfect foresight regarding their solar PV generation and demand, allowing them to optimize their strategy effectively. In reality, prosumers would face uncertainties in forecasting their solar PV output and consumption, which would affect their dispatch strategies and potentially lead to a less flexible behavior. This limitation appears to be particularly relevant for the real-time pricing scenario, where prosumer batteries are discharged in a way that minimizes grud consumption costs.\\

Finally, the analysis is conducted under current regulatory and policy frameworks. Future changes in policies affecting feed-in tariffs, prosumer self-consumption benefits, and grid tariff design could substantially change prosumage investment decision and behavior. For example, lower volumetric charges could substantially reduce the incentives for solar prosumage \cite{gunther_prosumage_2021}.

\section{Conclusion}
By linking capacity expansion, dispatch and grid models, we systematically assess the impact of tariff designs on the investment choices and operational strategies of solar prosumers, as well as their subsequent effects on the transmission grid. Within our model parameterization, findings reveal that investments into rooftop solar PV and home batteries are substantially driven by average solar availability, which is higher in the South of Germany. Zonal pricing provides stronger incentives for higher PV and battery investments compared to nodal pricing. The highest investments occur under zonal real-time pricing, which enables prosumers to align their self-consumption strategies better with the wholesale electricity market.\\

Regarding dispatch decisions, real-time pricing leads to more dynamic prosumer behavior, as battery storage is strategically used to minimize grid consumption costs by discharging when electricity prices are high. This results in a smoother overall demand curve. The broader electricity market effects however remain limited in our parameterization, with only minor differences in annual electricity generation across different technologies. The impact of prosumage on the transmission grid is similarly limited in our model setup. Network congestion is predominantly driven by exchang with other market areas rather than by prosumers. While small redispatch measures are observed, they do not significantly differ between tariff schemes. This suggests that under the conditions examined in this study, prosumer dispatch strategies only have a minor effect on overall transmission grid congestion.\\

This finding, however, could change if prosumers used their PV-batteries also for enery arbitrage with grid electricity, guided by wholesale prices. In a single price zone, charging and discharging of batteries could then even increase transmission bottlenecks, depending on the location of the batteries. Exploring this in an adjusted model setting could be a promising avenue for future research.
Likewise, our results hinge on several assumptions and simplifications of our modeling approach. Future research could address some of these limitations, including the anticipation of prosumer behavior in market-clearing mechanisms, the integration of additional household flexibility options such as electric vehicles and heat pumps, and the consideration of imperfect foresight in prosumer decision-making.

\section{Acknowledgments}
We gratefully acknowledge financial support by the German Federal Ministry of Economic Affairs and Climate Action (BMWK) via the project MODEZEEN (FKZ 03EI1019D). We thank Christoph Weyhing and Richard Weinhold for their contributions in the early stages of this work. 

\section{Author contributions}
\textbf{Dana Kirchem:} Methodology, Software, Formal analysis, Investigation, Data curation,  Writing - Original Draft, Visualization. \textbf{Mario Kendziorski:} Methodology, Software, Formal analysis, investigation, Data curation,  Writing - Original Draft, Visualization. \textbf{Enno Wiebrow:} Methodology, Software, Formal analysis, Investigation. \textbf{Wolf-Peter Schill:} Conceptualization, Methodology, Investigation, Writing - Review \& Editing, Funding acquisition. \textbf{Claudia Kemfert:} Writing - Review. \textbf{Christian von Hirschhausen:} Conceptualization, Funding acquisition.

\nomenclature[P]{$c^{curt}$}{Curtailment cost in EUR/MWh}
\nomenclature[P]{$mc_{p,t}$}{Marginal cost of generating unit $p$ at time $t$ in EUR/MWh}
\nomenclature[P]{$mc_{s,t}$}{Marginal cost of storage unit $s$ at time $t$ in EUR/MWh}
\nomenclature[P]{$load_{n,t}$}{Load at node $n$ at time $t$ in MWh}
\nomenclature[P]{$demand_{t}$}{Household electricity demand at time $t$ in MWh}
\nomenclature[P]{$c^{redisp}$}{Redispatch cost in EUR/MWh}
\nomenclature[P]{$c^{invest}_{pv}$}{Investment cost of rooftop PV in EUR/MWh}
\nomenclature[P]{$c^{fix}_{pv}$}{Fixed cost of rooftop PV in EUR/MWh}
\nomenclature[P]{$c^{invest}_{s^e}$}{Investment cost for home battery storage in EUR/MWh (energy)}
\nomenclature[P]{$c^{invest}_{s^p}$}{Investment cost for home battery storage in EUR/MW (power rating)}
\nomenclature[P]{$c^{fix}_{s}$}{Fixed cost for home battery storage in EUR/MWh}
\nomenclature[P]{$t^{fix}$}{Fixed share of volumetric retail tariff in EUR/MWh}
\nomenclature[P]{$t^{var}$}{Variable share of volumetric retail tariff in EUR/MWh}
\nomenclature[P]{$t^{feed}$}{Feed-in tariff in EUR/MWh}
\nomenclature[P]{$binary^{nodal}$}{Binary variable to indicate nodal real-time pricing [0,1]}
\nomenclature[P]{$binary^{zonal}$}{Binary variable to indicate zonal real-time pricing [0,1}
\nomenclature[P]{$rtp^{nodal}_t$}{Nodal real-time electricity price in EUR/MWh}
\nomenclature[P]{$rtp^{zonal}_t$}{Zonal real-time electricity price in EUR/MWh}

\nomenclature[S]{$N$}{Set of nodes}
\nomenclature[S]{$P$}{Set of power plants (including renewables)}
\nomenclature[S]{$S$}{Set of storage units}
\nomenclature[S]{$T$}{Set of time periods}
\nomenclature[S]{$Z$}{Set of market zones}

\nomenclature[V]{$CHARGE_{s,t}$}{Charging of storage unit $s$ at time $t$ in MW}
\nomenclature[V]{$CHARGE_{s,t}^{down}$}{Charging decrease of storage unit $s$ at time $t$ after redispatch in MW}
\nomenclature[V]{$CHARGE_{s,t}^{redisp}$}{Charging of storage unit $s$ at time $t$ after redispatch in MW}
\nomenclature[V]{$CHARGE_{s,t}^{up}$}{Charging increase of storage unit $s$ at time $t$ after redispatch in MW}
\nomenclature[V]{$CU_{p,t}^{redisp}$}{Curtailment of generating unit $p$ at time $t$ after redispatch in MWh}
\nomenclature[V]{$CU_{z,t}$}{Curtailment at zone $z$ at time $t$ in MWh}
\nomenclature[V]{$EX_{z,t}^{net}$}{Net exchange at zone $z$ at time $t$ in MWh}
\nomenclature[V]{$GEN_{p,t}$}{Generation of unit $p$ at time $t$ in MWh}
\nomenclature[V]{$GEN^{grid}_{pv,t}$}{Generation of rooftop PV to the grid at time $t$ in MWh}
\nomenclature[V]{$GEN^{self}_{pv,t}$}{Generation of rooftop PV for self-consumption at time $t$ in MWh}
\nomenclature[V]{$GEN_{p,t}^{redisp}$}{Generation of unit $p$ at time $t$ after redispatch in MW}
\nomenclature[V]{$INJ_{n,t}$}{Injection at node $n$ at time $t$ in MW}
\nomenclature[V]{$INVEST_{pv}$}{Investment of rooftop PV in MW}
\nomenclature[V]{$INVEST_{s^e}$}{Investment of home battery storage in MWh (energy)}
\nomenclature[V]{$INVEST_{s^p}$}{Investment of home battery storage in MW (power rating)}
\nomenclature[V]{$F_t$}{Grid consumption in MWh}

\newpage
\printnomenclature

 \bibliographystyle{elsarticle-num} 
 \bibliography{references}





\end{document}